\documentclass[aps,twocolumn,pra,tightenlines]{revtex4}
\usepackage{amsmath,amsthm,amsfonts,amssymb,amsbsy,mathrsfs}
\usepackage{units,times,float,graphicx,bm,comment}
\usepackage[colorlinks,citecolor=blue,linkcolor=blue,hyperindex,dvipdfm]{hyperref}
\begin{document}
\title{Quantum speed limit based on the bound of Bures angle}
\author{Shao-xiong Wu$^1$\footnote{sxwu@nuc.edu.cn}}
\author{Chang-shui Yu$^2$\footnote{ycs@dlut.edu.cn}}
\affiliation{$^1$ School of Science, North University of China,
Taiyuan 030051, China\\
$^2$ School of Physics, Dalian University of Technology, Dalian 116024, China}
\date{\today }

\begin{abstract}
In this paper, we investigate the unified bound of quantum speed limit time in open systems based on the modified Bures angle. This bound is applied to the damped Jaynes-Cummings model and the dephasing model, and the analytical quantum speed limit time is obtained for both models. As an example,  the maximum coherent qubit state with white noise is chosen as the initial states for the damped Jaynes-Cummings model. It is found that the quantum speed limit time  in both the non-Markovian and the Markovian regimes can be decreased by the white noise compared with the pure state. In addition, for the dephasing model, we find that the quantum speed limit time is not only related to the coherence of initial state and non-Markovianity, but also dependent on the population of initial excited state.
\end{abstract}
\maketitle

\section{Introduction}
In the quantum information processing, the evolution of quantum systems are significant for both the closed and open systems. The quantum speed limit (QSL) time of the closed system is defined as the minimal evolution time (corresponding to the  maximal evolution velocity) from the initial state to its orthogonal state. A unified quantum speed limit time is given by the Mandelstan-Tamm (MT) bound and the Margolus-Levitin (ML) bound, i.e.,  $\tau_{\text{qsl}}=\max\{\pi\hbar/(2\Delta E),\pi\hbar/(2\langle E\rangle)\}$ \cite{Mandelstam45,Fleming73,Bhattacharyya83,Bekenstein81,Vaidman92,Margolus98,Anandan90,Frey16,Deffner17jpa,
Pati91,Levitin09}. The quantum speed limit is also related to other quantum information processing, such as the role of entanglement in QSL \cite{Giovannetti03}, the elementary derivation for passage time \cite{Brody03}, the geometric QSL  based on statistical distance\cite{Jones10,Braunstein94}, the quantum evolution control\cite{Caneva09}, the relationship among with coherence and asymmetry \cite{Marvian16}, and so on.

In the practical scenarios, due to the interaction with surroundings, the evolution of quantum system should be treated with open system theory \cite{Breuer02}. Recently, the concepts of quantum speed limit were extended to the open quantum systems. For example, Taddei \emph{et al}. investigated the QSL employing the quantum Fisher information \cite{Taddei13} through the method developed in the Ref. \cite{Escher11}. Using the relative purity, del Campo \emph{et al}. derived a MT type time-energy uncertainty relation \cite{Plenio13}. Utilizing the Bures angle, Deffner and Lutz arrived a unified QSL bound for initial pure state, and showed that non-Markovian effects could speed up the quantum evolution \cite{Deffner13}. Other forms of QSL in open system were also reported, such as the QSL in different environments \cite{Xu14,Liu15,Cai17,Dehdashti15,Xu18,Zhang16,Song17},  the initial-state dependence\cite{Wu15}, the geometric form for Wigner phase space\cite{Deffner17njp}, the experimentally realizable metric \cite{Mondal16}. In addition, many other aspects of QSL were also widely studied such as using the fidelity \cite{Ektesabi17,Sun15}  and function of relative purity \cite{Zhang14,Wu18}, the mechanism for quantum speedup \cite{Liu16}, the connection with generation of quantumness \cite{Jing16}, generalization of geometric QSL form \cite{Pires16}, via gauge invariant distance \cite{Sun19}, even the QSL for almost all states \cite{Campaioli18}, and so on.

As a measure of distance, the Bures angle based on the Uhlmann fidelity has good properties, such as contractivity and triangle inequality. And, it is applied to the field of quantum speed limit in recently\cite{Deffner13}. However, the Bures angle is hard to measure the quantum speed limit for initial mixed state because it needs to calculate the square roots of matrices \cite{Braunstein94}. In the Ref. \cite{Miszczak09,Mendonca}, the authors derived an upper bound of Uhlmann fidelity (modified fidelity) between the mixed states, and obtained the upper bound of Bures angle. In this paper, we obtained the bound of quantum speed limit time for the initial mixed state according to the upper bound of Bures angle. The results showed that this bound is always tighter than the bound based on the Bures angle. For two-level system, the modified fidelity is consistent with the Uhlmann fidelity. So, the bound of the quantum speed limit based on the modified Bures angle is tight. As an application, this bound is employed to the damped Jaynes-Cummings model and dephasing model, respectively. The quantum speed limit time for both models are obtained analytically. As an example with generality, the maximum coherent qubit state with white noise is chosen as the initial state for the damped Jaynes-Cummings model. The evolution of the quantum system can be accelerated not only in the non-Markovian regime but also in the Markovian regime, and the quantum speed limit time will become short with the increasing of white noise. While, for the dephasing model, the quantum speed limit time is not only related to the coherence of initial state and non-Markovianity, but also dependent on the population of initial excited state. Generally speaking, the quantum speed limit is affected by many factors (such as the structure of environment, the form of the initial state), and the comprehensive competition of them determine the properties of quantum speed limit time.

\section{Results}

In the quantum information processing, the Bures angle $\mathcal{L}(\rho,\sigma)=\arccos[\sqrt{F(\rho,\sigma)}]$ is commonly used to measure the distance between the states $\rho$ and $\sigma$ with the Uhlmann fidelity $F(\rho,\sigma)=(\text{tr}[\sqrt{\sqrt{\rho}\sigma\sqrt{\rho}}])^{2}$. In the field of quantum speed limit, Bures angle is employed to the initial pure state \cite{Deffner13}, where the Bures angle can be simplified as $\mathcal{L}(\rho_{0},\rho_{t})=\arccos[\sqrt{\langle\psi_{0}\vert\rho_{t}\vert\psi_{0}\rangle}]$. However, due to calculation of the square roots of matrices, it is hard to obtain the quantum speed limit time in open system for the initial mixed state. Utilizing the relative purity\cite{Zhang14,Wu18}, the quantum speed limit can be extend to the initial mixed state, however relative purity and its function is not an optimal distance metric even for two-level system in some cases (similar  numerical simulation \cite{Wu14}). In the Refs.\cite{Miszczak09,Mendonca}, an upper bound of Uhlmann fidelity between mixed states and the modified Bures angle are proposed. Employing this modified Bures angle, we give a unified bound of quantum speed limit time, which is tight for initial two-level state or pure state.

The  upper bound of  Uhlmann fidelity $\mathcal{F}(\rho,\sigma)$ and the Uhlman fidelity $F(\rho,\sigma)$ satisfy the  inequality $F(\rho,\sigma)\leqslant\mathcal{F}(\rho,\sigma)$ \cite{Miszczak09,Mendonca}, where $\mathcal{F}(\rho,\sigma)$ is defined as
\begin{align}
\mathcal{F}(\rho,\sigma) & =\text{tr}[\rho\sigma]+\sqrt{1-\text{tr}[\rho^{2}]}\sqrt{1-\text{tr}[\sigma^{2}]}. \label{f2}
\end{align}
The modified Bures angle is defined as
\begin{align}
\Theta(\rho,\sigma)=\arccos[\sqrt{\mathcal{F}(\rho,\sigma)}],
\end{align}
and it meets the following inequality with the Bures angle
\begin{align}
\Theta(\rho_{0},\rho_{t})\leqslant\mathcal{L}(\rho_{0},\rho_{t}).
\end{align}

Using the derivation in the Method section, we can have a unified bound of the quantum speed limit time
\begin{eqnarray}
\tau_{\text{qsl}}=\max\left\{ \frac{1}{\varLambda_{\tau}^{\text{op}}},\frac{1}{\varLambda_{\tau}^{\text{tr}}},\frac{1}{\varLambda_{\tau}^{\text{hs}}}\right\} \sin^{2}[\Theta(\rho_{0},\rho_{\tau})],\label{eq4}
\end{eqnarray}
where
\begin{align}
\varLambda_{\tau}^{\text{op}} & =\frac{1}{\tau}\int_{0}^{\tau}dt\|L_{t}(\rho_{t})\|_{\text{op}}\left(1+\sqrt{\frac{1-\text{tr}[\rho_{0}^{2}]}{1-\text{tr}[\rho_{t}^{2}]}}\right),\nonumber \\
\varLambda_{\tau}^{\text{tr}} & =\frac{1}{\tau}\int_{0}^{\tau}dt\|L_{t}(\rho_{t})\|_{\text{tr}}\left(1+\sqrt{\frac{1-\text{tr}[\rho_{0}^{2}]}{1-\text{tr}[\rho_{t}^{2}]}}\right)
\label{eq25i}
\end{align}
and
\begin{align}
\varLambda_{\tau}^{\text{hs}}=\frac{1}{\tau}\int_{0}^{\tau}dt\|L_{t}(\rho_{t})\|_{\text{hs}}\left(1+\sqrt{\frac{1-\text{tr}[\rho_{0}^{2}]}{1-\text{tr}[\rho_{t}^{2}]}}\right).
\label{eq29i}
\end{align}

 According to the relationship among the norm of matrix $\|A\|_{\text{tr}}\geqslant\|A\|_{\text{hs}}\geqslant\|A\|_{\text{op}}$ \cite{Horn85}, the ``velocity" of quantum evolution satisfies the inequality $\varLambda_{\tau}^{\text{op}}\leqslant\varLambda_{\tau}^{\text{hs}}\leqslant\varLambda_{\tau}^{\text{tr}}$. Obviously, the ML bound based on operator norm provides the sharpest bound of quantum speed limit time in the open quantum system. As an application, it is applied to two paradigm models, i.e., the damped Jaynes-Cummings model and dephasing model.

\subsection{The damped Jaynes-Cummings model.} The total Hamiltonian of system and reservoir is $H=\frac{1}{2}\omega_{0}\sigma_{z}+\sum_{k}\omega_{k}b_{k}^{\dagger}b_{k}+\sum_{k}\left(g_{k}\sigma_{+}b_{k}+\text{h.c}\right)$,
and the evolution of reduced system is described by the master equation
\begin{align}
L_{t}(\rho_{t})=\frac{\gamma_{t}}{2}\left(2\sigma_{-}\rho_{t}\sigma_{+}-\sigma_{+}\sigma_{-}\rho_{t}-\rho_{t}\sigma_{+}\sigma_{-}\right),
\end{align}
where $\gamma_{t}$ is the time-dependent decay rate. The quantum system at time $\tau$ is analytically given by
\begin{align}
\rho({\tau})=\left(\begin{array}{cc}
\rho_{11}(0)\vert q_{\tau}\vert^{2} & \rho_{10}(0)q_{\tau}\\
\rho_{01}(0)q_{\tau}^{*} & 1-\rho_{11}(0)\vert q_{\tau}\vert^{2}
\end{array}\right)
\end{align}
with parameter $q_{\tau}=e^{-\Gamma_{\tau}/2}$, $\Gamma_{\tau}=\int_{0}^{\tau}dt\gamma_{t}$. Without loss of
generality, assuming the structure of non-Markovian reservoir is Lorentzian form
\begin{align}
J(\omega)=\frac{\gamma_{0}}{2\pi}\frac{\lambda^{2}}{(\omega_{0}-\omega)^{2}+\lambda^{2}},
\end{align}
where $\lambda$ is the spectral width of reservoir and $\gamma_{0}$
is the coupling strength between the system and reservoir. The ratio
$\gamma_{0}/\lambda$ determines the non-Markovianity of quantum dynamics. When $\gamma_{0}/\lambda>1/2$,
non-Markovian effect can influence the evolution of system distinctly\cite{Breuer02}.
Time-dependent decay rate $\gamma_{t}$ and parameter $q_{\tau}$
can be given with the explicit form as
\begin{align}
\gamma_{t} & =\frac{2\gamma_{0}\lambda\sinh(ht/2)}{h\cosh(ht/2)+\lambda\sinh(ht/2)},\nonumber \\
q_{\tau} & =e^{-\frac{\lambda\tau}{2}}\left[\cosh\left(\frac{h\tau}{2}\right)+\frac{\lambda}{h}\sinh\left(\frac{h\tau}{2}\right)\right]
\end{align}
with parameter $h=\sqrt{\lambda^{2}-2\gamma_{0}\lambda}$.

Turning into the Bloch representation, the mixed initial state can be expressed as
\begin{align}
\rho(0)=\left(
          \begin{array}{cc}
            \rho_{11}(0) & \rho_{10}(0) \\
            \rho_{01}(0) & 1-\rho_{11}(0) \\
          \end{array}
        \right)=\frac{1}{2}\left(
                  \begin{array}{cc}
                    1+r_z & r_x-ir_y \\
                    r_x+ir_y & 1-r_z \\
                  \end{array}
                \right),\label{initialstate}
\end{align}
where, $r_x,r_y,r_z$ are the Bloch vectors. The quantum speed limit time for the mixed initial state (\ref{initialstate}) is
\begin{align}
\tau_{\text{qsl}}=\frac{1+r_z-q_t(r_x^2+r_y^2+q_{\tau}r_z(1+r_z))-\kappa_1\kappa_2^{\tau}} {\frac{1}{\tau}\int_0^{\tau}dt\vert\dot{q}_t\sqrt{r_x^2+r_y^2+4q_t^2(1+r_z)^2}(1+\frac{\kappa_1}{\kappa_2^t})\vert},
\label{jc-qsl}
\end{align}
where the parameters $\kappa_1=\sqrt{1-r_x^2-r_y^2-r_z^2}$ and $\kappa_2^t=\sqrt{q_t^2(2+2r_z-r_x^2-r_y^2-q_t^2(1+r_z)^2)}$. For the two-level quantum state (\ref{initialstate}), one can follow the Ref \cite{Wu18}, and investigate the effect of coherence of the initial state and the population of initial excited state on the quantum speed limit time.

As an example with generality, we will assume the initial state to be a  two-level maximally coherent state $\vert\psi\rangle=(\vert0\rangle+\vert1\rangle)/\sqrt{2}$ with white noise
\begin{align}
\rho(0)=\frac{1-p}{2}\mathbb{I}+p\vert\psi\rangle\langle\psi\vert,\label{werner}
\end{align}
where $\mathbb{I}$ (identity matrix) means the white noise, and $p\in[0,1]$ is the component of $\vert\psi\rangle$.
The tightest ML bound of quantum speed limit time can be given analytically as
\begin{align}
\tau_{\text{qsl}}=\frac{1-p^{2}q_{\tau}-\kappa_{1\text{w}}\kappa_{2\text{w}}^{\tau}}
{\frac{1}{\tau}\int_{0}^{\tau}dt\vert\sqrt{p^{2}+4q_{t}^{2}}\dot{q}_{t}
(1+\frac{\kappa_{1\text{w}}}{\kappa_{2\text{w}}^t})\vert} \label{eq24}
\end{align}
with $\kappa_{1\text{w}}=\sqrt{1-p^{2}}$ and $\kappa_{2\text{w}}^t=\sqrt{q_{t}^{2}(2-p^{2}-q_{t}^{2})}$.
One can find that the quantum speed limit time (\ref{eq24}) is determined by the white noise and the interaction with the environment. In Fig. 1, we show the ratio between the quantum speed limit time and actual driven time $\tau_{\text{qsl}}/\tau$ for the initial state (\ref{werner}) as functions of the coupling strength $\gamma_0$ and the component of white noise, which is expressed as $1-p$. The actual driven time is $\tau=1$ and the non-Markovian parameter is chosen as $\lambda=15$ (in unit of $\omega_0$). As the previous results in the Ref. \cite{Wu15,Wu18}, the evolution of the system will be accelerated not only in the non-Markovian regime but also in the Markovian regime when the initial state is not the excited state. And, we can observe that the quantum speed limit time reaches the maximum when $\gamma_0$ is in the vicinity of $\lambda/2$, and becomes shorter as the increasing of white noise. From the perspective that the quantum state will evolve to a full mixed state when the time is enough long, a reasonable explanation is that the quantum state with large purity will change more significantly when the initial state is pure and the evolution time is finite, and the discrimination between the initial and final state can be measured using fidelity or Bures angle. So, the quantum speed limit time will be shorter when the component of the white noise is larger.

\begin{figure}
\begin{center}
  \includegraphics[width=0.7\columnwidth]{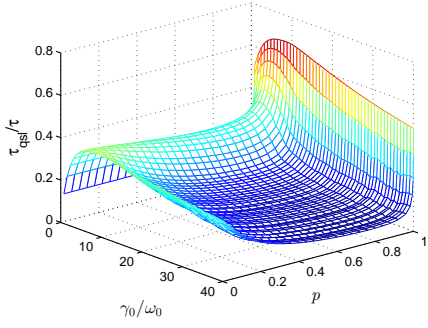}
   \caption{\textbf{ The ratio between the quantum speed limit time and actual driven time $\tau_{\text{qsl}}/\tau$ of qubit state (\ref{initialstate}) for damped Jaynes-Cumming model.} The spectral width parameter is chosen as $\lambda=15$ (in unit of $\omega_{0}$), and the actual driving time is $\tau=1$. }
  \end{center}
\end{figure}

When the initial state is  maximum coherence state $\vert\psi\rangle=(\vert1\rangle+\vert0\rangle)/\sqrt{2}$, i.e., without white noise, the quantum speed limit time (\ref{eq24}) can be simplified as
\begin{eqnarray}
\tau_{\text{qsl}}=\frac{1-q_{\tau}}{\frac{1}{\tau}\int_{0}^{\tau}dt\vert\dot{q}_{t}\sqrt{1+4q_{t}^{2}}\vert},\label{JCpure}
\end{eqnarray}
which agrees with the result reported in the Ref. \cite{Wu15} based on the Bures angle $\mathcal{L}(\rho,\sigma)$.

\subsection{The dephasing model.} It can be described as spin-boson form interaction between qubit system and a bosonic reservoir, the total Hamiltonian is $H=\frac{1}{2}\omega_{0}\sigma_{z}+\sum_{k}\omega_{k}b_{k}^{\dagger}b_{k}+\sum_{k}\sigma_{z}(g_{k}b_{k}^{\dagger}+g_{k}^{*}b_{k})$.
The dynamics of reduced quantum system $\rho_{t}$ is
\begin{align}
L_{t}(\rho_{t})=\frac{\gamma_{t}}{2}\left(\sigma_{z}\rho_{t}\sigma_{z}-\rho_{t}\right).
\end{align}
For the initial state in Bloch representation (\ref{initialstate}), the reduced state in time $\tau$ has the following form
\begin{align}
\rho({\tau})=\frac{1}{2}\left(\begin{array}{cc}
1+r_z & (r_x-ir_y)e^{-\Gamma_{\tau}}\\
 (r_x+ir_y)e^{-\Gamma_{\tau}} & 1-r_z
\end{array}\right).
\end{align}
Taking the continuum limit of reservoir mode and assuming the spectrum of reservoir $J(\omega)$, the dephasing factor $\Gamma_{\tau}$ can be given explicitly as \cite{Breuer02}
\begin{eqnarray}
\Gamma_{\tau}=\int_{0}^{\infty}\text{d}\omega J(\omega)\coth\left(\frac{\omega}{2k_{B}T}\right)\frac{1-\cos\omega\tau}{\omega^{2}},
\end{eqnarray}
where $k_{B}$ is the Boltzmann's constant and $T$ is temperature.

For the zero temperature condition, choosing Ohmic-like spectrum with soft cutoff $J(\omega)=\eta\omega^{s}/\omega_{c}^{s-1}\exp\left(-\omega/\omega_{c}\right)$, and assuming the cutoff frequency $\omega_{c}$ is unit, the dephasing factor $\Gamma_{\tau}$ can be solved analytically as \cite{Chin13}
\begin{eqnarray}
\Gamma_{\tau}=\eta\left[1-\frac{\cos[(s-1)\arctan\tau]}{(1+\tau^{2})^{(s-1)/2}}\right]\Gamma(s-1),
\end{eqnarray}
where $\Gamma(\cdot)$ is the Euler gamma function and $\eta$ is
dimensionless constant. The property of environment is determined
by parameter $s$, and the reservoir can be divided into the sub-Ohmic
reservoir ($s<1$), Ohmic reservoir ($s=1$) and super-Ohmic reservoir
($s>1$). The dephasing rate $\gamma_t$, i.e., the derivative of dephasing factor $\Gamma_{t}$, has analytical form $\gamma_{t}=\eta(1+t^{2})^{-s/2}\Gamma(s)\sin[s\arctan t]$.

The ML bound of quantum speed limit time based on the operator norm can be given as
\begin{align}
\tau_{\text{qsl}}=\frac{1-\mathcal{C}^2e^{-\Gamma_{\tau}}-\langle\sigma_z\rangle^2-\chi_1\chi_2^{\tau}} {\frac{1}{\tau}\int_0^{\tau}dt\vert\gamma_t\mathcal{C} e^{-\Gamma_t}(1+\frac{\chi_1}{\chi_2^{\tau}})\vert},
\label{qsldephasing1}
\end{align}
where the parameters $\chi_1=\sqrt{1-\mathcal{C}^2-\langle\sigma_z\rangle^2}$ and $\chi_1^t=\sqrt{1-\mathcal{C}^2e^{-2\Gamma_t}-\langle\sigma_z\rangle^2}$. In the Eq. (\ref{qsldephasing1}), we use the fact that the coherence of initial state (\ref{initialstate}) satisfied $\mathcal{C}^2=r_x^2+r_y^2$ and the Bloch vector $r_z$ means the population of initial excited state $\langle\sigma_z\rangle$.

\begin{figure}
\begin{center}
  \includegraphics[width=0.7\columnwidth]{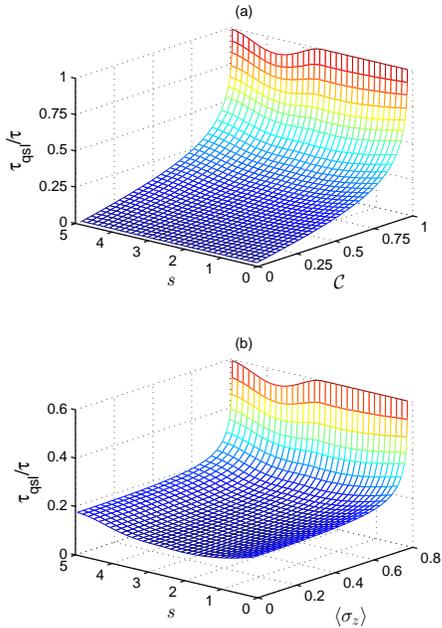}
  \caption{\textbf{The ratio between quantum speed limit time and actual driven time $\tau_{\text{qsl}}/\tau$  for the  dephasing model.} (a) The ratio $\tau_{\text{qsl}}/\tau$ is the functions of the Ohmic parameter $s$ and the coherence of initial state $\mathcal{C}$. The $\langle\sigma_z\rangle$ is chosen as zero. (b) The ratio $\tau_{\text{qsl}}/\tau$ varies as  with the Ohmic parameter $s$ and $\langle\sigma_z\rangle$. The coherence of initial state is $\mathcal{C}=0.6$. In both the panels (a) and (b), the actual driven time are chosen as constant $\tau=3$.}
  \end{center}
\end{figure}

In Fig. 2(a), we demonstrate the ratio between the quantum speed limit time (\ref{qsldephasing1}) and the actual driven time $\tau_{\text{qsl}}/\tau$ as functions of the Ohmic parameter $s$ and coherence of initial state $\mathcal{C}$. The actual driven time is constant $\tau=3$ and $\langle\sigma_z\rangle$ is chosen as zero. One can observe that the bound of quantum speed limit time will be tighter when the coherence of initial state $\mathcal{C}$ become greater. Compared with the quantum coherence, the effect of non-Markoviantity (corresponded to $\gamma_t$ and related to $s$) on the quantum speed limit time is weaker. In the non-Markovian regime, the quantum speed limit time will decrease slightly. The physical analysis of similar phenomenon for pure initial state based on the Bures angle is given in the previous Ref. \cite{Wu15}. In Fig. 2(b), we show the ratio between the quantum speed limit time (\ref{qsldephasing1}) and the actual driven time $\tau_{\text{qsl}}/\tau$ as functions of the Ohmic parameter $s$ and the population of initial excited state $\langle\sigma_z\rangle$. The actual driven time is chosen as constant $\tau=3$ and the coherence of initial state is $\mathcal{C}=0.6$ . It is easy to find that the quantum speed limit time is influenced strongly by the population $\langle\sigma_z\rangle$ and increases rapidly as the population $\langle\sigma_z\rangle$  become larger.

One can observe that the quantum speed limit time (\ref{qsldephasing1}) is not only related to the coherence of initial state and the non-Markovianity of dynamics, but also dependent on the population of initial excited state. It is different from the results using the function of relative purity \cite{Zhang14,Wu18}, where the quantum speed limit time is independent of $\langle\sigma_z\rangle$. For a mixed initial state, the dephasing processing means that losing of information without losing of energy, so the energy of the system (related to $\langle\sigma_z\rangle$) influences the system evolution is reasonable and physical consistent. So, the quantum speed limit time (\ref{qsldephasing1}) recovers more information about the dephasing processing.

\section{Discussion}

The quantum speed limit play important roles in both the closed and open systems, and the experiment implementation in the cavity QED has been reported \cite{qed15}. Utilizing the upper bound of Uhlmann fidelity, we investigated the unified bound of quantum speed limit time in open systems based on the modified Bures angle, and this bound is tight for pure state and qubit state. We applied this bound to the damped Jaynes-Cummings model and dephasing model, and obtained the analytical results for both models. For the damped Jaynes-Cummings model, the maximum coherent qubit state with white noise is chosen as the initial state, and its quantum speed limit time can be decreased not only in the non-Markovian regime but also in the Markovian regime, and can be influenced significantly by even small noises. While, for the dephasing model, the quantum speed limit time is not only related to the coherence of initial state and non-Markovianity, but also dependent on the population of initial excited state. It should be noted the bound of quantum speed limit time (\ref{eq4}) maybe fail to measure the evolution of high dimensional mixed system, and the general quantum speed limit of mixed quantum system deserves further investigation.

\section{Method}

In this section, we will derive the quantum speed limit of open quantum systems. Consider the time derivative of modified Bures angle $\Theta$,
\begin{align}
\frac{d}{dt}\Theta(\rho_{0},\rho_{t}) & \leqslant\vert\frac{d}{dt}\Theta(\rho_{0},\rho_{t})\vert\nonumber \\
 & =\frac{\vert\dot{\mathcal{F}}(\rho_{0},\rho_{t})\vert}{2\sqrt{1-\mathcal{F}(\rho_{0},\rho_{t})}\sqrt{\mathcal{F}(\rho_{0},\rho_{t})}},\label{eq5}
\end{align}
where the time derivative of modified fidelity $\mathcal{F}(\rho_{0},\rho_{t})$ in Eq. \eqref{f2} is given as follows:
\begin{align}
\dot{\mathcal{F}}(\rho_{0},\rho_{t}) & =\vert\text{tr}[\rho_{0}\dot{\rho}_{t}]-\sqrt{\frac{1-\text{tr}[\rho_{0}^{2}]}{1-\text{tr}[\rho_{t}^{2}]}}\text{tr}[\rho_{t}\dot{\rho}_{t}]\vert\nonumber \\
 & \leqslant\left\vert \text{tr}[\rho_{0}\dot{\rho}_{t}]\right\vert +\sqrt{\frac{1-\text{tr}[\rho_{0}^{2}]}{1-\text{tr}[\rho_{t}^{2}]}}\left\vert \text{tr}[\rho_{t}\dot{\rho}_{t}]\right\vert.
\end{align}
When the dynamics of quantum systems is non-unitary, the evolution
of quantum state is expressed by $\dot{\rho}_{t}=L_{t}(\rho_{t})$.
Substituting the definition of $\Theta(\rho_{0},\rho_{t})$ into Eq.
(\ref{eq5}), the derivative of modified Bures angle $\Theta$ can be rewritten
as
\begin{align}
2\cos[\Theta]\sin[\Theta]\dot{\Theta}\leqslant\left\vert \text{tr}[\rho_{0}L_{t}(\rho_{t})]\right\vert +\sqrt{\frac{1-\text{tr}[\rho_{0}^{2}]}{1-\text{tr}[\rho_{t}^{2}]}}\left\vert \text{tr}[\rho_{t}L_{t}(\rho_{t})]\right\vert .\label{eq8}
\end{align}
For any $n\times n$ complex  matrices $A_{1}$ and $A_{2}$, there is von
Neumann inequality
\begin{equation}
\left\vert \text{tr}[A_{1}A_{2}]\right\vert \leqslant\sum_{i=1}^{n}\sigma_{1,i}\sigma_{2,i}
\end{equation}
with the descending singular values $\sigma_{1,1}\geqslant\cdots\geqslant\sigma_{1,n}$
and $\sigma_{2,1}\geqslant\cdots\geqslant\sigma_{2,n}$. For the first
item of right side in Eq. (\ref{eq8}), one can have
\begin{align}
\left|\text{tr}[\rho_{0}L_{t}(\rho_{t})]\right|\leqslant\sum_{i}p_{i}\lambda_{i},
\end{align}
where $p_{i}$ are the singular values of state $\rho_{0}$, and
$\lambda_{i}$ are the singular values of operator $L_{t}(\rho_{t})$.
For the second item in Eq. (\ref{eq8}), we can obtain that
\begin{align}
\left\vert \text{tr}[\rho_{t}L_{t}(\rho_{t})]\right\vert \leqslant\sum_{i}\epsilon_{i}\lambda_{i},
\end{align}
where $\epsilon_{i}$ are the singular values of state $\rho_{t}$.

Since $p_{i}\leqslant1$ and $\epsilon_{i}\leqslant1$, one can obtain
that $\sum_{i}p_{i}\lambda_{i}\leqslant\lambda_{1}\leqslant\sum_{i}\lambda_{i}$
and $\sum_{i}\epsilon_{i}\lambda_{i}\leqslant\lambda_{1}\leqslant\sum_{i}\lambda_{i}$.
For operator $L_{t}(\rho_{t})$, the largest singular value $\lambda_{1}$
can be expressed as operator norm $\|L_{t}(\rho_{t})\|_{\text{op}}$
and the sum of $\lambda_{i}$ can be expressed as trace norm $\|L_{t}(\rho_{t})\|_{\text{tr}}$.

Similar to the Ref. \cite{Deffner13}, the Margolus-Levitin bound of quantum speed limit time of open system can be given
by
\begin{eqnarray}
\tau_{\text{qsl}}=\max\left\{ \frac{1}{\varLambda_{\tau}^{\text{op}}},\frac{1}{\varLambda_{\tau}^{\text{tr}}}\right\} \sin^{2}[\Theta(\rho_{0},\rho_{\tau})],\label{qslml}
\end{eqnarray}
where the denominators in the above equation are defined as
\begin{align}
\varLambda_{\tau}^{\text{op}} & =\frac{1}{\tau}\int_{0}^{\tau}dt\|L_{t}(\rho_{t})\|_{\text{op}}\left(1+\sqrt{\frac{1-\text{tr}[\rho_{0}^{2}]}{1-\text{tr}[\rho_{t}^{2}]}}\right),\nonumber \\
\varLambda_{\tau}^{\text{tr}} & =\frac{1}{\tau}\int_{0}^{\tau}dt\|L_{t}(\rho_{t})\|_{\text{tr}}\left(1+\sqrt{\frac{1-\text{tr}[\rho_{0}^{2}]}{1-\text{tr}[\rho_{t}^{2}]}}\right).
\label{eq25}
\end{align}

Applying the Cauchy-Schwarz inequality for operators, i.e., $\vert\text{tr}[A_{1}^{\dag}A_{2}]\vert^{2}\leqslant\text{tr}[A_{1}^{\dag}A_{1}]\text{tr}[A_{2}^{\dag}A_{2}]$,
the Eq. (\ref{eq8}) can be rewritten as
\begin{align}
 & 2\cos[\Theta]\sin[\Theta]\dot{\Theta}\nonumber \\
\leqslant & \sqrt{\text{tr}[\rho_{0}^{2}]}\sqrt{\text{tr}[L_{t}^{\dag}(\rho_{t})L_{t}(\rho_{t})]}\nonumber \\
 & +\sqrt{\frac{1-\text{tr}[\rho_{0}^{2}]}{1-\text{tr}[\rho_{t}^{2}]}}\sqrt{\text{tr}[\rho_{t}^{2}]}\sqrt{\text{tr}[L_{t}^{\dag}(\rho_{t})L_{t}(\rho_{t})]}\nonumber \\
\leqslant & \left(1+\sqrt{\frac{1-\text{tr}[\rho_{0}^{2}]}{1-\text{tr}[\rho_{t}^{2}]}}\right)\sqrt{\text{tr}[L_{t}^{\dag}(\rho_{t})L_{t}(\rho_{t})]}.
\end{align}
The fact that the purity of density matrix satisfies $\text{tr}[\rho^{2}]\leqslant1$
for both states $\rho_{0}$ and $\rho_{t}$ is used in the last inequality. And, $\sqrt{\text{tr}[L_{t}^{\dag}(\rho_{t})L_{t}(\rho_{t})]}$ is the
Hilbert-Schmidt norm of operator $L_{t}(\rho_{t})$, which is
defined as $\|L_{t}(\rho_{t})\|_{\text{hs}}=\sqrt{\sum_{i}\lambda_{i}^{2}}$.
So, the Eq. (\ref{eq8}) can be simplified as
\begin{align}
2\cos[\Theta]\sin[\Theta]\dot{\Theta}\leqslant\left(1+\sqrt{\frac{1-\text{tr}[\rho_{0}^{2}]}{1-\text{tr}[\rho_{t}^{2}]}}\right)\|L_{t}(\rho_{t})\|_{\text{hs}}.
\end{align}
So, the Mandelstam-Tamm bound quantum speed limit time of non-unitary dynamics $L_{t}(\rho_{t})$ is
\begin{align}
\tau_{\text{qsl}}=\frac{1}{\varLambda_{\tau}^{\text{hs}}}\sin^{2}[\Theta(\rho_{0},\rho_{\tau})],\label{eq16}
\end{align}
where
\begin{align}
\varLambda_{\tau}^{\text{hs}}=\frac{1}{\tau}\int_{0}^{\tau}dt\|L_{t}(\rho_{t})\|_{\text{hs}}\left(1+\sqrt{\frac{1-\text{tr}[\rho_{0}^{2}]}{1-\text{tr}[\rho_{t}^{2}]}}\right).
\label{eq29}
\end{align}
Combining the Eqs. (\ref{qslml}) and (\ref{eq16}), the unified expression
of quantum speed limit time based on the modified Bures angle for initial mixed state is given by
\begin{eqnarray*}
\tau_{\text{qsl}}=\max\left\{ \frac{1}{\varLambda_{\tau}^{\text{op}}},\frac{1}{\varLambda_{\tau}^{\text{tr}}},\frac{1}{\varLambda_{\tau}^{\text{hs}}}\right\} \sin^{2}[\Theta(\rho_{0},\rho_{\tau})].
\end{eqnarray*}

\section*{Acknowledgments}

Wu was supported by the Scientific and Technological Innovation Programs of Higher Education Institutions in Shanxi under Grant No. 2019L0527. Yu was supported by the National Natural Science Foundation of China under Grant No.11775040, and the Fundamental Research Fund for the Central Universities under Grant No. DUT18LK45.

\end{document}